 \documentclass[conference,10pt,twoside,twocolumn]{IEEEtran}

\usepackage{mleftright}
\usepackage{cite}
\usepackage[T1]{fontenc}
\usepackage{graphicx}
\usepackage{mathtools}
\usepackage{amssymb}
\usepackage{amsmath}
\usepackage{paralist}
\usepackage{subfigure}
\usepackage{booktabs} 
\usepackage{algorithm}
\usepackage{algorithmic}
\usepackage{amsthm}
\usepackage{multirow}
\usepackage{microtype}

\usepackage{amssymb}
\usepackage{amsfonts}
\usepackage{mathrsfs}
\usepackage{xspace}
\usepackage{bm}
\usepackage{upgreek}

\newcommand{\safemath}[2]{\newcommand{#1}{\ensuremath{#2}\xspace}}

\safemath{\bma}{\mathbf{a}}
\safemath{\bmb}{\mathbf{b}}
\safemath{\bmc}{\mathbf{c}}
\safemath{\bmd}{\mathbf{d}}
\safemath{\bme}{\mathbf{e}}
\safemath{\bmf}{\mathbf{f}}
\safemath{\bmg}{\mathbf{g}}
\safemath{\bmh}{\mathbf{h}}
\safemath{\bmi}{\mathbf{i}}
\safemath{\bmj}{\mathbf{j}}
\safemath{\bmk}{\mathbf{k}}
\safemath{\bml}{\mathbf{l}}
\safemath{\bmm}{\mathbf{m}}
\safemath{\bmn}{\mathbf{n}}
\safemath{\bmo}{\mathbf{o}}
\safemath{\bmp}{\mathbf{p}}
\safemath{\bmq}{\mathbf{q}}
\safemath{\bmr}{\mathbf{r}}
\safemath{\bms}{\mathbf{s}}
\safemath{\bmt}{\mathbf{t}}
\safemath{\bmu}{\mathbf{u}}
\safemath{\bmv}{\mathbf{v}}
\safemath{\bmw}{\mathbf{w}}
\safemath{\bmx}{\mathbf{x}}
\safemath{\bmy}{\mathbf{y}}
\safemath{\bmz}{\mathbf{z}}
\safemath{\bmzero}{\mathbf{0}}
\safemath{\bmone}{\mathbf{1}}

\bmdefine{\biad}{a}
\bmdefine{\bibd}{b}
\bmdefine{\bicd}{c}
\bmdefine{\bidd}{d}
\bmdefine{\bied}{e}
\bmdefine{\bifd}{f}
\bmdefine{\bigd}{g}
\bmdefine{\bihd}{h}
\bmdefine{\biid}{i}
\bmdefine{\bijd}{j}
\bmdefine{\bikd}{k}
\bmdefine{\bild}{l}
\bmdefine{\bimd}{m}
\bmdefine{\bind}{n}
\bmdefine{\biod}{o}
\bmdefine{\bipd}{p}
\bmdefine{\biqd}{q}
\bmdefine{\bird}{r}
\bmdefine{\bisd}{s}
\bmdefine{\bitd}{t}
\bmdefine{\biud}{u}
\bmdefine{\bivd}{v}
\bmdefine{\biwd}{w}
\bmdefine{\bixd}{x}
\bmdefine{\biyd}{y}
\bmdefine{\bizd}{z}

\bmdefine{\bixid}{\xi}
\bmdefine{\bilambdad}{\lambda}
\bmdefine{\bimud}{\mu}
\bmdefine{\bithetad}{\theta}
\bmdefine{\biphid}{\phi}
\bmdefine{\bideltad}{\delta}

\safemath{\bmia}{\biad}
\safemath{\bmib}{\bibd}
\safemath{\bmic}{\bicd}
\safemath{\bmid}{\bidd}
\safemath{\bmie}{\bied}
\safemath{\bmif}{\bifd}
\safemath{\bmig}{\bigd}
\safemath{\bmih}{\bihd}
\safemath{\bmii}{\biid}
\safemath{\bmij}{\bijd}
\safemath{\bmik}{\bikd}
\safemath{\bmil}{\bild}
\safemath{\bmim}{\bimd}
\safemath{\bmin}{\bind}
\safemath{\bmio}{\biod}
\safemath{\bmip}{\bipd}
\safemath{\bmiq}{\biqd}
\safemath{\bmir}{\bird}
\safemath{\bmis}{\bisd}
\safemath{\bmit}{\bitd}
\safemath{\bmiu}{\biud}
\safemath{\bmiv}{\bivd}
\safemath{\bmiw}{\biwd}
\safemath{\bmix}{\bixd}
\safemath{\bmiy}{\biyd}
\safemath{\bmiz}{\bizd}

\safemath{\bmxi}{\bixid}
\safemath{\bmlambda}{\bilambdad}
\safemath{\bmmu}{\bimud}
\safemath{\bmtheta}{\bithetad}
\safemath{\bmphi}{\biphid}
\safemath{\bmdelta}{\bideltad}

\safemath{\bA}{\mathbf{A}}
\safemath{\bB}{\mathbf{B}}
\safemath{\bC}{\mathbf{C}}
\safemath{\bD}{\mathbf{D}}
\safemath{\bE}{\mathbf{E}}
\safemath{\bF}{\mathbf{F}}
\safemath{\bG}{\mathbf{G}}
\safemath{\bH}{\mathbf{H}}
\safemath{\bI}{\mathbf{I}}
\safemath{\bJ}{\mathbf{J}}
\safemath{\bK}{\mathbf{K}}
\safemath{\bL}{\mathbf{L}}
\safemath{\bM}{\mathbf{M}}
\safemath{\bN}{\mathbf{N}}
\safemath{\bO}{\mathbf{O}}
\safemath{\bP}{\mathbf{P}}
\safemath{\bQ}{\mathbf{Q}}
\safemath{\bR}{\mathbf{R}}
\safemath{\bS}{\mathbf{S}}
\safemath{\bT}{\mathbf{T}}
\safemath{\bU}{\mathbf{U}}
\safemath{\bV}{\mathbf{V}}
\safemath{\bW}{\mathbf{W}}
\safemath{\bX}{\mathbf{X}}
\safemath{\bY}{\mathbf{Y}}
\safemath{\bZ}{\mathbf{Z}}

\safemath{\bZero}{\mathbf{0}}
\safemath{\bOne}{\mathbf{1}}
\safemath{\bDelta}{\mathbf{\Delta}}
\safemath{\bLambda}{\mathbf{\UpLambda}}
\safemath{\bPhi}{\mathbf{\Upphi}}
\safemath{\bSigma}{\mathbf{\Upsigma}}
\safemath{\bOmega}{\mathbf{\Upomega}}
\safemath{\bTheta}{\mathbf{\Uptheta}}

\bmdefine{\biAd}{A}
\bmdefine{\biBd}{B}
\bmdefine{\biCd}{C}
\bmdefine{\biDd}{D}
\bmdefine{\biEd}{E}
\bmdefine{\biFd}{F}
\bmdefine{\biGd}{G}
\bmdefine{\biHd}{H}
\bmdefine{\biId}{I}
\bmdefine{\biJd}{J}
\bmdefine{\biKd}{K}
\bmdefine{\biLd}{L}
\bmdefine{\biMd}{M}
\bmdefine{\biOd}{N}
\bmdefine{\biPd}{O}
\bmdefine{\biQd}{P}
\bmdefine{\biRd}{R}
\bmdefine{\biSd}{S}
\bmdefine{\biTd}{T}
\bmdefine{\biUd}{U}
\bmdefine{\biVd}{V}
\bmdefine{\biWd}{W}
\bmdefine{\biXd}{X}
\bmdefine{\biYd}{Y}
\bmdefine{\biZd}{Z}

\bmdefine{\biDelta}{\Delta}
\bmdefine{\biLambda}{\Lambda}
\bmdefine{\biPhi}{\Phi}
\bmdefine{\biSigma}{\Sigma}
\bmdefine{\biOmega}{\Omega}
\bmdefine{\biTheta}{\Theta}

\safemath{\bimA}{\biAd}
\safemath{\bimB}{\biBd}
\safemath{\bimC}{\biCd}
\safemath{\bimD}{\biDd}
\safemath{\bimE}{\biEd}
\safemath{\bimF}{\biFd}
\safemath{\bimG}{\biGd}
\safemath{\bimH}{\biHd}
\safemath{\bimI}{\biId}
\safemath{\bimJ}{\biJd}
\safemath{\bimK}{\biKd}
\safemath{\bimL}{\biLd}
\safemath{\bimM}{\biMd}
\safemath{\bimN}{\biNd}
\safemath{\bimO}{\biOd}
\safemath{\bimP}{\biPd}
\safemath{\bimQ}{\biQd}
\safemath{\bimR}{\biRd}
\safemath{\bimS}{\biSd}
\safemath{\bimT}{\biTd}
\safemath{\bimU}{\biUd}
\safemath{\bimV}{\biVd}
\safemath{\bimW}{\biWd}
\safemath{\bimX}{\biXd}
\safemath{\bimY}{\biYd}
\safemath{\bimZ}{\biZd}

\safemath{\bimDelta}{\biDelta}
\safemath{\bimLambda}{\biLambda}
\safemath{\bimPhi}{\biPhi}
\safemath{\bimSigma}{\biSigma}
\safemath{\bimOmega}{\biOmega}
\safemath{\bimTheta}{\biTheta}

\safemath{\setA}{\mathcal{A}}
\safemath{\setB}{\mathcal{B}}
\safemath{\setC}{\mathcal{C}}
\safemath{\setD}{\mathcal{D}}
\safemath{\setE}{\mathcal{E}}
\safemath{\setF}{\mathcal{F}}
\safemath{\setG}{\mathcal{G}}
\safemath{\setH}{\mathcal{H}}
\safemath{\setI}{\mathcal{I}}
\safemath{\setJ}{\mathcal{J}}
\safemath{\setK}{\mathcal{K}}
\safemath{\setL}{\mathcal{L}}
\safemath{\setM}{\mathcal{M}}
\safemath{\setN}{\mathcal{N}}
\safemath{\setO}{\mathcal{O}}
\safemath{\setP}{\mathcal{P}}
\safemath{\setQ}{\mathcal{Q}}
\safemath{\setR}{\mathcal{R}}
\safemath{\setS}{\mathcal{S}}
\safemath{\setT}{\mathcal{T}}
\safemath{\setU}{\mathcal{U}}
\safemath{\setV}{\mathcal{V}}
\safemath{\setW}{\mathcal{W}}
\safemath{\setX}{\mathcal{X}}
\safemath{\setY}{\mathcal{Y}}
\safemath{\setZ}{\mathcal{Z}}
\safemath{\emptySet}{\varnothing}

\safemath{\colA}{\mathscr{A}}
\safemath{\colB}{\mathscr{B}}
\safemath{\colC}{\mathscr{C}}
\safemath{\colD}{\mathscr{D}}
\safemath{\colE}{\mathscr{E}}
\safemath{\colF}{\mathscr{F}}
\safemath{\colG}{\mathscr{G}}
\safemath{\colH}{\mathscr{H}}
\safemath{\colI}{\mathscr{I}}
\safemath{\colJ}{\mathscr{J}}
\safemath{\colK}{\mathscr{K}}
\safemath{\colL}{\mathscr{L}}
\safemath{\colM}{\mathscr{M}}
\safemath{\colN}{\mathscr{N}}
\safemath{\colO}{\mathscr{O}}
\safemath{\colP}{\mathscr{P}}
\safemath{\colQ}{\mathscr{Q}}
\safemath{\colR}{\mathscr{R}}
\safemath{\colS}{\mathscr{S}}
\safemath{\colT}{\mathscr{T}}
\safemath{\colU}{\mathscr{U}}
\safemath{\colV}{\mathscr{V}}
\safemath{\colW}{\mathscr{W}}
\safemath{\colX}{\mathscr{X}}
\safemath{\colY}{\mathscr{Y}}
\safemath{\colZ}{\mathscr{Z}}

\safemath{\opA}{\mathbb{A}}
\safemath{\opB}{\mathbb{B}}
\safemath{\opC}{\mathbb{C}}
\safemath{\opD}{\mathbb{D}}
\safemath{\opE}{\mathbb{E}}
\safemath{\opF}{\mathbb{F}}
\safemath{\opG}{\mathbb{G}}
\safemath{\opH}{\mathbb{H}}
\safemath{\opI}{\mathbb{I}}
\safemath{\opJ}{\mathbb{J}}
\safemath{\opK}{\mathbb{K}}
\safemath{\opL}{\mathbb{L}}
\safemath{\opM}{\mathbb{M}}
\safemath{\opN}{\mathbb{N}}
\safemath{\opO}{\mathbb{O}}
\safemath{\opP}{\mathbb{P}}
\safemath{\opQ}{\mathbb{Q}}
\safemath{\opR}{\mathbb{R}}
\safemath{\opS}{\mathbb{S}}
\safemath{\opT}{\mathbb{T}}
\safemath{\opU}{\mathbb{U}}
\safemath{\opV}{\mathbb{V}}
\safemath{\opW}{\mathbb{W}}
\safemath{\opX}{\mathbb{X}}
\safemath{\opY}{\mathbb{Y}}
\safemath{\opZ}{\mathbb{Z}}
\safemath{\opZero}{\mathbb{O}}
\safemath{\identityop}{\opI}

\safemath{\veca}{\bma}
\safemath{\vecb}{\bmb}
\safemath{\vecc}{\bmc}
\safemath{\vecd}{\bmd}
\safemath{\vece}{\bme}
\safemath{\vecf}{\bmf}
\safemath{\vecg}{\bmg}
\safemath{\vech}{\bmh}
\safemath{\veci}{\bmi}
\safemath{\vecj}{\bmj}
\safemath{\veck}{\bmk}
\safemath{\vecl}{\bml}
\safemath{\vecm}{\bmm}
\safemath{\vecn}{\bmn}
\safemath{\veco}{\bmo}
\safemath{\vecp}{\bmp}
\safemath{\vecq}{\bmq}
\safemath{\vecr}{\bmr}
\safemath{\vecs}{\bms}
\safemath{\vect}{\bmt}
\safemath{\vecu}{\bmu}
\safemath{\vecv}{\bmv}
\safemath{\vecw}{\bmw}
\safemath{\vecx}{\bmx}
\safemath{\vecy}{\bmy}
\safemath{\vecz}{\bmz}

\safemath{\veczero}{\bmzero}
\safemath{\vecone}{\bmone}
\safemath{\vecxi}{\bmxi}
\safemath{\veclambda}{\bmlambda}
\safemath{\vecmu}{\bmmu}
\safemath{\vectheta}{\bmtheta}
\safemath{\vecphi}{\bmphi}
\safemath{\vecdelta}{\bmdelta}

\safemath{\matA}{\bA}
\safemath{\matB}{\bB}
\safemath{\matC}{\bC}
\safemath{\matD}{\bD}
\safemath{\matE}{\bE}
\safemath{\matF}{\bF}
\safemath{\matG}{\bG}
\safemath{\matH}{\bH}
\safemath{\matI}{\bI}
\safemath{\matJ}{\bJ}
\safemath{\matK}{\bK}
\safemath{\matL}{\bL}
\safemath{\matM}{\bM}
\safemath{\matN}{\bN}
\safemath{\matO}{\bO}
\safemath{\matP}{\bP}
\safemath{\matQ}{\bQ}
\safemath{\matR}{\bR}
\safemath{\matS}{\bS}
\safemath{\matT}{\bT}
\safemath{\matU}{\bU}
\safemath{\matV}{\bV}
\safemath{\matW}{\bW}
\safemath{\matX}{\bX}
\safemath{\matY}{\bY}
\safemath{\matZ}{\bZ}
\safemath{\matzero}{\bmzero}

\safemath{\matDelta}{\bDelta}
\safemath{\matLambda}{\bLambda}
\safemath{\matPhi}{\bPhi}
\safemath{\matSigma}{\bSigma}
\safemath{\matOmega}{\bOmega}
\safemath{\matTheta}{\bTheta}

\safemath{\matidentity}{\matI}
\safemath{\matone}{\matO}

\safemath{\rnda}{A}
\safemath{\rndb}{B}
\safemath{\rndc}{C}
\safemath{\rndd}{D}
\safemath{\rnde}{E}
\safemath{\rndf}{F}
\safemath{\rndg}{G}
\safemath{\rndh}{H}
\safemath{\rndi}{I}
\safemath{\rndj}{J}
\safemath{\rndk}{K}
\safemath{\rndl}{L}
\safemath{\rndm}{M}
\safemath{\rndn}{N}
\safemath{\rndo}{O}
\safemath{\rndp}{P}
\safemath{\rndq}{Q}
\safemath{\rndr}{R}
\safemath{\rnds}{S}
\safemath{\rndt}{T}
\safemath{\rndu}{U}
\safemath{\rndv}{V}
\safemath{\rndw}{W}
\safemath{\rndx}{X}
\safemath{\rndy}{Y}
\safemath{\rndz}{Z}

\safemath{\rveca}{\bimA}
\safemath{\rvecb}{\bimB}
\safemath{\rvecc}{\bimC}
\safemath{\rvecd}{\bimD}
\safemath{\rvece}{\bimE}
\safemath{\rvecf}{\bimF}
\safemath{\rvecg}{\bimG}
\safemath{\rvech}{\bimH}
\safemath{\rveci}{\bimI}
\safemath{\rvecj}{\bimJ}
\safemath{\rveck}{\bimK}
\safemath{\rvecl}{\bimL}
\safemath{\rvecm}{\bimM}
\safemath{\rvecn}{\bimN}
\safemath{\rveco}{\bomO}
\safemath{\rvecp}{\bimP}
\safemath{\rvecq}{\bimQ}
\safemath{\rvecr}{\bimR}
\safemath{\rvecs}{\bimS}
\safemath{\rvect}{\bimT}
\safemath{\rvecu}{\bimU}
\safemath{\rvecv}{\bimV}
\safemath{\rvecw}{\bimW}
\safemath{\rvecx}{\bimX}
\safemath{\rvecy}{\bimY}
\safemath{\rvecz}{\bimZ}

\safemath{\rvecxi}{\bmxi}
\safemath{\rveclambda}{\bmlambda}
\safemath{\rvecmu}{\bmmu}
\safemath{\rvectheta}{\bmtheta}
\safemath{\rvecphi}{\bmphi}

\safemath{\rmatA}{\bimA}
\safemath{\rmatB}{\bimB}
\safemath{\rmatC}{\bimC}
\safemath{\rmatD}{\bimD}
\safemath{\rmatE}{\bimE}
\safemath{\rmatF}{\bimF}
\safemath{\rmatG}{\bimG}
\safemath{\rmatH}{\bimH}
\safemath{\rmatI}{\bimI}
\safemath{\rmatJ}{\bimJ}
\safemath{\rmatK}{\bimK}
\safemath{\rmatL}{\bimL}
\safemath{\rmatM}{\bimM}
\safemath{\rmatN}{\bimN}
\safemath{\rmatO}{\bimO}
\safemath{\rmatP}{\bimP}
\safemath{\rmatQ}{\bimQ}
\safemath{\rmatR}{\bimR}
\safemath{\rmatS}{\bimS}
\safemath{\rmatT}{\bimT}
\safemath{\rmatU}{\bimU}
\safemath{\rmatV}{\bimV}
\safemath{\rmatW}{\bimW}
\safemath{\rmatX}{\bimX}
\safemath{\rmatY}{\bimY}
\safemath{\rmatZ}{\bimZ}

\safemath{\rmatDelta}{\bimDelta}
\safemath{\rmatLambda}{\bimLambda}
\safemath{\rmatPhi}{\bimPhi}
\safemath{\rmatSigma}{\bimSigma}
\safemath{\rmatOmega}{\bimOmega}
\safemath{\rmatTheta}{\bimTheta}

\usepackage{amssymb}
\usepackage{amsfonts}
\usepackage{mathrsfs}
\usepackage{xspace}
\usepackage{bm}
\usepackage{fancyref}
\usepackage{textcomp}

\usepackage{multirow}
\usepackage{stmaryrd}

\newenvironment{textbmatrix}{	\setlength{\arraycolsep}{2.5pt}%
								\big[\begin{matrix}}{\end{matrix}\big]%
								\raisebox{0.08ex}{\vphantom{M}}}

\def\be{\begin{equation}}
\def\ee{\end{equation}}
\def\een{\nonumber \end{equation}}
\def\mat{\begin{bmatrix}}
\def\emat{\end{bmatrix}}
\def\btm{\begin{textbmatrix}}
\def\etm{\end{textbmatrix}}

\def\ba#1\ea{\begin{align}#1\end{align}}
\def\bas#1\eas{\begin{align*}#1\end{align*}}
\def\bs#1\es{\begin{split}#1\end{split}}
\def\bg#1\eg{\begin{gather}#1\end{gather}}
\def\bml#1\eml{\begin{multline}#1\end{multline}}
\def\bi#1\ei{\begin{itemize}#1\end{itemize}}

\newcommand{\lefto}{\mathopen{}\left}

\DeclareMathOperator{\Exop}{\opE}			
\DeclareMathOperator{\Varop}{\opV\!\mathrm{ar}} 

\newcommand{\Ex}[2]{\ensuremath{\Exop_{#1}\lefto[#2\right]}} 	
\newcommand{\abs}[1]{\lefto\lvert#1\right\rvert}		

\safemath{\dirac}{\delta}					
\safemath{\krond}{\dirac}					

\safemath{\upto}{\uparrow}
\safemath{\downto}{\downarrow}
\safemath{\iu}{j}							
\safemath{\ev}{\lambda}						
\safemath{\hilseqspace}{l^{2}}				
\newcommand{\banachfunspace}[1]{\setL^{#1}}	
\safemath{\hilfunspace}{\banachfunspace{2}}	

\safemath{\SNR}{\textit{SNR}} 				
\safemath{\PAR}{\textit{PAR}} 				
\safemath{\No}{N_0}							
\safemath{\Es}{E_s}							
\safemath{\Eb}{E_b}							
\safemath{\EbNo}{\frac{\Eb}{\No}}
\safemath{\EsNo}{\frac{\Es}{\No}}

\DeclareMathOperator{\CHop}{\ensuremath{\opH}} 
\safemath{\tvir}{\rndh_{\CHop}}				
\safemath{\tvtf}{\rndl_{\CHop}}				
\safemath{\spf}{\rnds_{\CHop}}				
\safemath{\bff}{H_{\CHop}}					

\safemath{\ircf}{r_{h}}						
\safemath{\tftvcf}{r_{s}}					
\safemath{\tfcf}{r_{l}}						
\safemath{\bfcf}{r_{H}}						

\safemath{\tcorr}{c_h}						
\safemath{\scf}{c_{s}}						
\safemath{\tfcorr}{c_{l}}					
\safemath{\fcorr}{c_{H}}						

\safemath{\mi}{I}							
\safemath{\capacity}{C}						

\safemath{\normal}{\mathcal{N}}			
\safemath{\jpg}{\mathcal{CN}}			
\safemath{\mchain}{\leftrightarrow}		

\safemath{\dB}{\,\mathrm{dB}}
\safemath{\dBm}{\,\mathrm{dBm}}
\safemath{\Hz}{\,\mathrm{Hz}}
\safemath{\kHz}{\,\mathrm{kHz}}
\safemath{\MHz}{\,\mathrm{MHz}}
\safemath{\GHz}{\,\mathrm{GHz}}
\safemath{\s}{\,\mathrm{s}}
\safemath{\ms}{\,\mathrm{ms}}
\safemath{\mus}{\,\mathrm{\text{\textmu}s}}
\safemath{\ns}{\,\mathrm{ns}}
\safemath{\ps}{\,\mathrm{ps}}
\safemath{\meter}{\,\mathrm{m}}
\safemath{\mm}{\,\mathrm{mm}}
\safemath{\cm}{\,\mathrm{cm}}
\safemath{\m}{\,\mathrm{m}}
\safemath{\W}{\,\mathrm{W}}
\safemath{\mW}{\, \mathrm{mW}}
\safemath{\J}{\,\mathrm{J}}
\safemath{\K}{\,\mathrm{K}}
\safemath{\bit}{\,\mathrm{bit}}
\safemath{\nat}{\,\mathrm{nat}}

\safemath{\define}{\triangleq}			

\safemath{\equivalent}{\sim}
\safemath{\distas}{\sim}					
\safemath{\sdiff}{\Delta}				

\safemath{\reals}{\mathbb{R}}
\safemath{\positivereals}{\reals_{+}}
\safemath{\integers}{\mathbb{Z}}
\safemath{\posint}{\integers_{+}}
\safemath{\naturals}{\mathbb{N}}
\safemath{\posnaturals}{\naturals_{+}}
\safemath{\complexset}{\mathbb{C}}
\safemath{\rationals}{\mathbb{Q}}

\newcommand*{\fancyrefapplabelprefix}{app}		
\newcommand*{\fancyrefthmlabelprefix}{thm}		
\newcommand*{\fancyreflemlabelprefix}{lem}		
\newcommand*{\fancyrefcorlabelprefix}{cor}		
\newcommand*{\fancyrefdeflabelprefix}{def}		
\newcommand*{\fancyrefproplabelprefix}{prop}		
\newcommand*{\fancyrefexmpllabelprefix}{exmpl}
\newcommand*{\fancyrefalglabelprefix}{alg}		
\newcommand*{\fancyreftbllabelprefix}{tbl}		

\frefformat{vario}{\fancyrefseclabelprefix}{Section~#1}
\frefformat{vario}{\fancyrefthmlabelprefix}{Theorem~#1}
\frefformat{vario}{\fancyreftbllabelprefix}{Table~#1}
\frefformat{vario}{\fancyreflemlabelprefix}{Lemma~#1}
\frefformat{vario}{\fancyrefcorlabelprefix}{Corollary~#1}
\frefformat{vario}{\fancyrefdeflabelprefix}{Definition~#1}
\frefformat{vario}{\fancyreffiglabelprefix}{Fig.~#1}
\frefformat{vario}{\fancyrefapplabelprefix}{Appendix~#1}
\frefformat{vario}{\fancyrefeqlabelprefix}{(#1)}
\frefformat{vario}{\fancyrefproplabelprefix}{Proposition~#1}
\frefformat{vario}{\fancyrefexmpllabelprefix}{Example~#1}
\frefformat{vario}{\fancyrefalglabelprefix}{Algorithm~#1}

 \newtheorem{thm}{Theorem}
 
 \newtheorem{defi}{Definition}
 \newtheorem{lem}[thm]{Lemma}
 
\safemath{\dictab}{[\,\dicta\,\,\dictb\,]}

\safemath{\ysig}{\bmy}
\safemath{\ysighat}{\hat{\ysig}}
\safemath{\ysigdim}{M}
\safemath{\xsig}{\bmx}
\safemath{\xsigdim}{N}
\safemath{\nx}{n_x}
\safemath{\zsig}{\bmz}
\safemath{\zsigdim}{\ysigdim}
\safemath{\rsig}{\bmr}
\safemath{\Adict}{\bA}
\safemath{\Adicttilde}{\widetilde{\Adict}}
\safemath{\Adictdim}{\outputdim\times\xsigdim}
\safemath{\avec}{\bma}
\safemath{\avectilde}{\tilde{\avec}}
\safemath{\Bdict}{\bB}
\safemath{\Bdicttilde}{\widetilde{\Bdict}}
\safemath{\Cdict}{\bC}
\safemath{\cvec}{\bmc}
\safemath{\Ddict}{\bD}
\safemath{\Ddictdim}{\ysigdim\times\xsigdim}
\safemath{\dvec}{\bmd}
\safemath{\Ddicttilde}{\widetilde{\bD}}
\safemath{\Bonb}{\bB}
\safemath{\bvec}{\bmb}
\safemath{\Bonbdim}{\ysigdim\times\ysigdim}
\safemath{\noise}{\bmn}
\safemath{\noisedim}{\ysigim}
\safemath{\err}{\bme}
\safemath{\errdim}{\ysigdim}
\safemath{\errset}{\setE}
\safemath{\nerr}{n_e}
\safemath{\delop}{\bP_\errset}
\safemath{\delopc}{\bP_{{\errset}^c}}

\safemath{\cplxi}{\imath}
\safemath{\cplxj}{\jmath}

\safemath{\dict}{\matD}
\safemath{\inputdim}{N}		
\safemath{\outputdim}{M}		
\safemath{\sparsity}{S}	
\safemath{\inputdimA}{{N_a}}	
\safemath{\inputdimB}{{N_b}}	
\safemath{\elemA}{{n_a}}	
\safemath{\elemB}{{n_b}}	
\safemath{\resA}{\matR_a}	
\safemath{\resB}{\matR_b}	
\safemath{\subD}{\matS} 
\safemath{\subA}{\matS_a} 
\safemath{\subB}{\matS_b} 
\safemath{\dicta}{\matA} 	
\safemath{\dictb}{\matB} 	
\safemath{\hollowS}{H}
\safemath{\hollowA}{H_a}
\safemath{\hollowB}{H_b}
\safemath{\cross}{Z}
\safemath{\coh}{\mu_d}			
\safemath{\coha}{\mu_a}			
\safemath{\cohb}{\mu_b}			
\safemath{\mubs}{\nu}	
\safemath{\cohm}{\mu_m} 
\safemath{\dictset}{\setD}	
\safemath{\dictsetp}{\dictset(\coh,\coha,\cohb)}	
\safemath{\dictsetgen}{\dictset_\text{gen}}
\safemath{\dictsetgenp}{\dictsetgen(\coh)}
\safemath{\dictsetonb}{\dictset_\text{onb}}
\safemath{\dictsetonbp}{\dictsetonb(\coh)}

\safemath{\leftside}{U}
\safemath{\rightsideA}{R_a}
\safemath{\rightsideB}{R_b}

\safemath{\indexS}{\setI_S} 

\safemath{\na}{n_a}			
\safemath{\nb}{n_b}			
\safemath{\coeffa}{p_i}	
\safemath{\coeffb}{q_j}	
\safemath{\seta}{\setP}		
\safemath{\setb}{\setQ}     
\safemath{\setw}{\setW}	
\safemath{\setz}{\setZ}	
\safemath{\cola}{\veca}		
\safemath{\colb}{\vecb}		
\safemath{\cold}{\vecd}		
\safemath{\inputvec}{\vecx} 	
\safemath{\error}{\vece}	
\safemath{\noiseout}{\vecz} 	
\safemath{\inputvecel}{x}
\safemath{\inputveca}{\vecx_a}
\safemath{\inputvecb}{\vecx_b}
\safemath{\outputvec}{\vecy}	
\safemath{\lambdamin}{\lambda_{\mathrm{min}}}

\safemath{\elltwo}{\ell_2}
\safemath{\ellone}{\ell_1}
\safemath{\ellzero}{\ell_0}
\safemath{\ellinf}{\ell_\infty}
\safemath{\ellinftilde}{\ell_{\widetilde\infty}}
\safemath{\licard}{Z(\coh,\coha,\cohb)}
\safemath{\xsol}{\hat{x}}
\safemath{\xbord}{x_b}		
\safemath{\xstat}{x_s}		
\safemath{\xstatLone}{\tilde{x}_s}
\safemath{\order}{\mathcal{O}} 
\safemath{\scales}{\Theta} 
\safemath{\ones}{\mathbf{1}} 
\safemath{\zeroes}{\mathbf{0}} 
\safemath{\thlone}{\kappa(\coh,\cohb)} 
\safemath{\constoneA}{\delta} 
\safemath{\constoneB}{\epsilon} 
\safemath{\nlarge}{L}				   
\safemath{\sumlarge}{S_\nlarge}
\safemath{\maxlarger}{P_\nlarge}	   
\safemath{\Pzero}{\textrm{P0}}	
\safemath{\Pone}{\textrm{P1}}
\safemath{\vecfir}{\vecw}			 
\safemath{\vecsec}{\vecz}
\safemath{\elvecfir}{w}              
\safemath{\elvecsec}{z}				 
\safemath{\nlargefir}{n}
\safemath{\normout}{\gamma}
\safemath{\auxfun}{h}
\safemath{\supp}{\textrm{supp}}

\safemath{\indexa}{\ell}
\safemath{\indexb}{r}
\safemath{\indexc}{i}
\safemath{\indexd}{j}

\safemath{\project}{P}

\newcommand{\dd}{\textnormal{d}}%

\usepackage{color}

\newcommand{\PD}{PD}
\newcommand{\PDspace}{PD }
\newcommand{\FD}{FD}

\safemath{\LAMA}{\textrm{LAMA}}

\safemath{\mLAMA}{\textrm{M-LAMA}}
\safemath{\smLAMA}{\textrm{SM-LAMA}}
\safemath{\mCBAMP}{\textrm{mcB-AMP}}

\safemath{\MRT}{\textrm{MRT}}
\safemath{\betamax}{\beta^\textnormal{max}}
\safemath{\tmax}{t_\textnormal{max}}
\safemath{\betamaxno}{\beta^\textnormal{max}}
\safemath{\betamin}{\beta^\textnormal{min}}
\safemath{\betaminno}{\beta^\textnormal{min}}

\safemath{\Nomin}{\No^\textnormal{min}(\beta)}
\safemath{\Nominnobeta}{\No^\textnormal{min}}
\safemath{\Nomax}{\No^\textnormal{max}(\beta)}
\safemath{\Nomaxnobeta}{\No^\textnormal{max}}

\safemath{\MAP}{\textrm{MAP}}
\safemath{\IO}{\textrm{IO}}
\safemath{\JO}{\textrm{JO}}
\safemath{\Nopost}{N_{0}^\textnormal{post}}
\safemath{\MT}{U}
\safemath{\MR}{B}
\safemath{\C}{C}
\safemath{\Tran}{\textnormal{T}}
\safemath{\bmymf}{\bmy^\textnormal{MF}}
\safemath{\bmymrc}{\bmy^\textnormal{MRC}}
\safemath{\Herm}{\textnormal{H}}
\safemath{\row}{\textnormal{r}}
\safemath{\col}{\textnormal{c}}

\IEEEoverridecommandlockouts 

\begin{document}

\title{On the Achievable Rates of Decentralized Equalization in Massive MU-MIMO Systems}

\author{Charles Jeon, Kaipeng Li, Joseph R. Cavallaro, and Christoph Studer
\thanks{CJ and CS are with the School of ECE at Cornell University, Ithaca, NY; e-mails: {jeon@csl.cornell.edu}, {studer@cornell.edu}.}
\thanks{KL and JRC are with the Department of ECE at Rice University, Houston, TX; e-mails: {kl33@rice.edu}, {cavallar@rice.edu}.}
\thanks{The work of CJ and CS was supported by Xilinx, Inc.~and by the US NSF under grants ECCS-1408006,  CCF-1420328, and CAREER CCF-1652065. The work of KL and JRC was supported by Xilinx, Inc.~and by the US NSF under grants CNS-1265332, ECCS-1232274, and ECCS-1408370.}
}

\maketitle

\begin{abstract}
%
Massive multi-user (MU) multiple-input multiple-output (MIMO) promises significant gains in spectral efficiency compared to traditional, small-scale MIMO technology. 
Linear equalization algorithms, such as zero forcing~(ZF) or minimum mean-square error (MMSE)-based methods, typically rely on centralized processing at the base station~(BS), which results in (i) excessively high interconnect and chip input/output data rates, and (ii) high computational complexity. 
In this paper, we investigate the achievable rates of decentralized equalization that mitigates both of these issues. 
We consider two distinct BS architectures that partition the antenna array into clusters, each associated with independent radio-frequency chains and signal processing hardware, and the results of each cluster are fused in a feedforward network. 
For both architectures, we consider ZF, MMSE, and a novel, non-linear equalization algorithm that builds upon approximate message passing (AMP), and we theoretically analyze the achievable rates of these methods. 
Our results demonstrate that decentralized equalization with our AMP-based methods incurs no or only a negligible loss in terms of achievable rates compared to that of centralized solutions. 
\end{abstract}



\section{Introduction}
\label{sec:intro}

Massive MU-MIMO is widely believed to be a key technology for next-generation wireless systems \cite{ABCHLAZ2014}.
By equipping the BS with hundreds or thousands of antennas and serving tens or hundreds of users simultaneously in the same time-frequency resource, massive MU-MIMO  enables orders-of-magnitude improvements in spectral efficiency compared to traditional, small-scale MIMO \cite{LETM2014}.
However, the presence of hundreds or thousands of antenna elements at the BS causes significant implementation challenges of this technology. 

One of the most critical implementation challenges is the excessively high amount of data that must be transferred between the BS antenna array and the baseband processing unit.
For example, the raw baseband data rates (from or to the RF chains) exceed $200$\,Gbit/s for a MU-MIMO system with $128$ BS antennas each using two $10$\,bit analog-to-digital converters operating with $40$\,MHz bandwidth.
Such high data rates cannot be sustained by existing interconnect technology and chip input/output (I/O) interfaces. Furthermore, existing conventional linear equalization algorithms, such as ZF and MMSE-based methods, rely on centralized processing and require excessively high computational complexity~\cite{LSCCGS2016}.
Hence, existing massive MU-MIMO testbeds either distribute baseband processing in the frequency domain~\cite{MVNKWOETL2016}, i.e., perform parallel computations per subcarrier, or use maximum ratio combining (MRC)~\cite{SYALMYZ2012}, which enables fully distributed processing at the antenna elements.
Frequency distribution, however, requires that each frequency cluster needs access to all the BS antennas, which prevents a scaling to thousands of antennas. MRC is known to result in poor spectral efficiency for realistic antenna configurations~\cite{HBD11}. 
Consequently, a practical deployment of massive MU-MIMO necessitates novel equalization methods that reduce the interconnect bandwidth and baseband processing complexity while maximizing the spectral efficiency.

\setlength{\textfloatsep}{10pt}
\begin{figure}[tp]
\centering
%
\includegraphics[width=0.71\columnwidth]{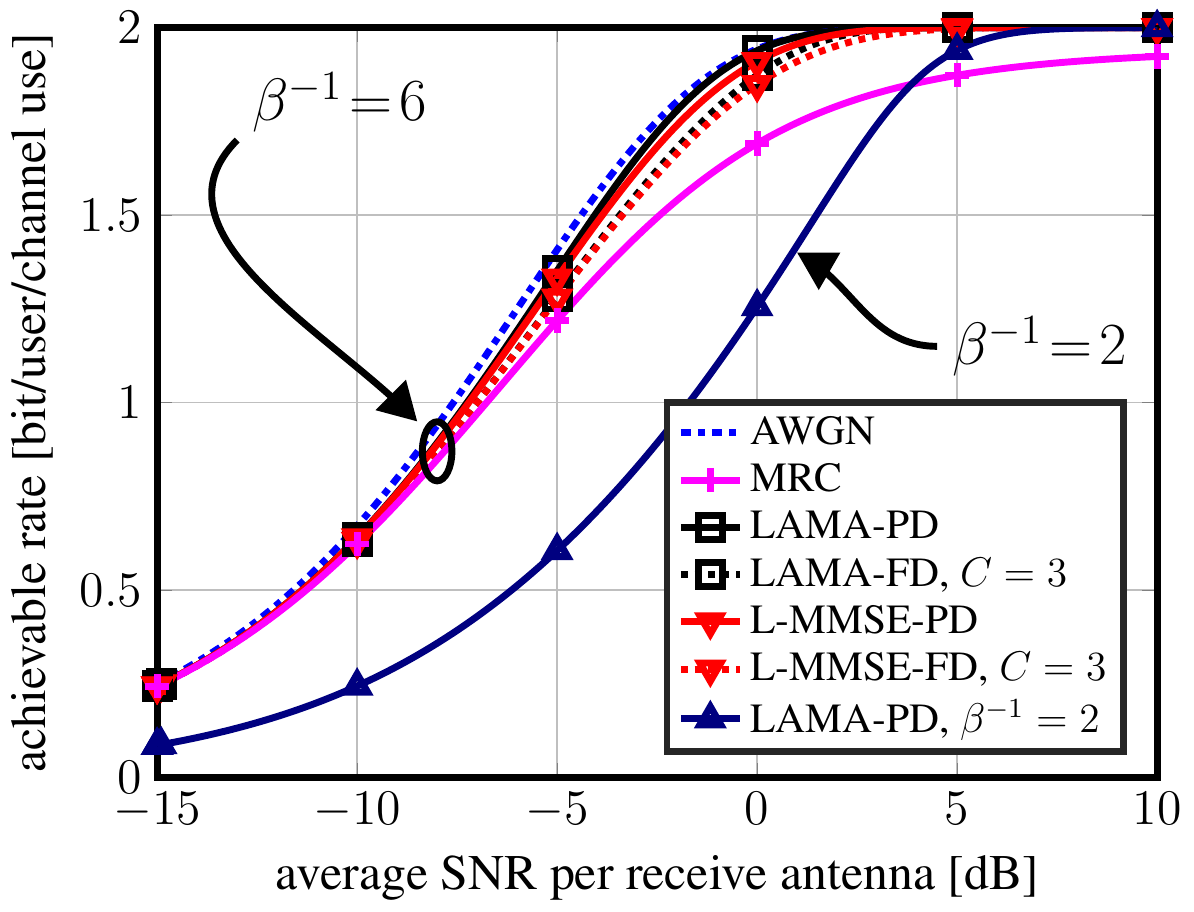}
\vspace{-0.2cm}
\caption{Achievable rates for QPSK in the large-system limit for $\beta^{-1}=\MR/\MT=6$.
%
Fully decentralized LAMA and L-MMSE with $C=3$ clusters only suffer a minimal rate loss compared to the fundamental AWGN performance limit.
%
MRC, which is decentralized by nature, performs poorly at higher rates. 
}
\label{fig:fixed_rate_1_99}
\end{figure}

\subsection{Contributions}
We propose two decentralized BS architectures, namely partially decentralized (PD) and fully decentralized (FD) equalization, which mitigate the interconnect, I/O, and computation bottlenecks.
For both of these architectures, we investigate the performance of ZF, linear MMSE (L-MMSE), and a novel, non-linear decentralized equalization method that builds upon our recently proposed \underline{la}rge \underline{M}IMO \underline{a}pproximate message passing (LAMA) algorithm~\cite{JGMS2015conf}.
We develop a state-evolution (SE) framework that enables a precise analysis of the achievable rates and error-rate performance of decentralized equalization in the large-system limit, and we show simulation results for realistic, finite-dimensional systems that agree with our theory.

\fref{fig:fixed_rate_1_99} demonstrates that FD equalization with $C=3$ antenna clusters in combination with LAMA achieves rates that are close to that of an interference-free AWGN channel even for moderate BS-to-user-antenna ratios. 
FD equalization achieves significantly higher rates than PD equalization with a lower BS-to-user-antenna ratio, which demonstrates that higher spectral efficiency can be achieved through decentralized architectures that reduce the interconnect and chip I/O bandwidths.

\subsection{Relevant prior art}

Architectures that perform decentralized processing in the spatial domain have been proposed in\mbox{\cite{LSCCGS2016,LCSGCS2016}}.\footnote{Distributed processing is also a key component of coordinated multipoint (CoMP) \cite{IDMGFBMTJ2011} and cloud radio access networks (CRANs) \cite{PLZW2015} for multi-cell transmission. The decentralized architectures and algorithms in~\cite{LSCCGS2016,LCSGCS2016} are specifically designed for massive MU-MIMO systems in which the computing hardware is collocated near the antenna array and within a single cell.}
The idea is to partition the BS antenna array into $C$ independent clusters, each associated with local computing hardware. Equalization and beamforming is then carried out in an iterative fashion by exchanging consensus information among the clusters. 
While these iterative methods significantly reduce the raw baseband data rates and the computation bottlenecks, their performance has not been analyzed and the throughput suffers from interconnect latency. 
In contrast, we focus on decentralized \emph{feedforward} architectures  whose performance can be analyzed theoretically and is less susceptible to interconnect latency.

%

%

One of the proposed decentralized equalization algorithms in this paper builds upon AMP \cite{donoho2009,BM2011}.
Centralized equalization via AMP was shown to achieve near individually-optimal (IO) performance for realistic massive MU-MIMO systems \cite{JGMS2015conf,JMS2016}.
A distributed version of AMP was proposed recently in~\cite{ZBB2016}. The key differences to this method are as follows: (i) we consider feedforward architectures, which are key for low-latency processing as required by next-generation massive MU-MIMO  systems~\cite{LETM2014}, and (ii) we analyze the achievable rates and error-rate performance in massive MU-MIMO systems. 

\subsection{Notation}
Lowercase and uppercase boldface letters designate vectors and matrices, respectively; uppercase calligraphic letters denote sets. The transpose and Hermitian of the matrix $\bH$ are represented by $\bH^\Tran$ and $\bH^\Herm$.
%
%
%
We define $\left\langle \bmx \right\rangle = \frac{1}{N}\sum_{k=1}^N x_k$. 
The multivariate complex-valued Gaussian probability density function (pdf) with mean $\bmm$ and covariance~$\bK$ is denoted by $\setC\setN(\bmm,\bK)$. $\Exop_X\!\left[\cdot\right]$ and $\Varop_X\!\left[\cdot\right]$ represent the mean and variance with respect to the 
random variable~$X$, respectively.


\begin{figure}[tp]
\centering
%
\subfigure[Partially decentralized (PD) equalization.]{\includegraphics[width=0.75\columnwidth]{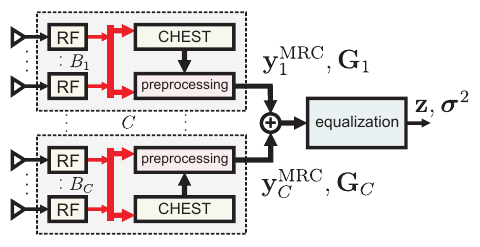}
\label{fig:pda}
}
\subfigure[Fully decentralized (FD) equalization.]{\includegraphics[width=0.75\columnwidth]{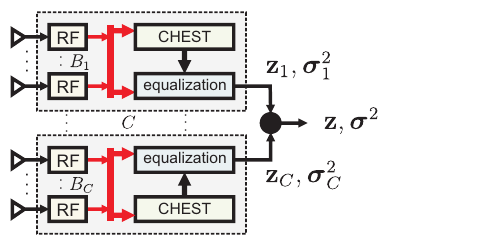}
\label{fig:fda}
}
\caption{Partially decentralized (PD) and fully decentralized (FD) equalization architectures for the massive MU-MIMO uplink with $\C$ clusters.
(a) PD performs decentralized channel estimation (CHEST) and preprocessing; equalization is performed in a centralized fashion and operates on low-dimensional data. (b) FD performs CHEST, preprocessing, and equalization in a decentralized manner. 
The $\oplus$ operator in (a) denotes matrix/vector-additions and $\bullet$ in (b) denotes a weighted vector addition (see \fref{sec:FDA} for the details).
}
\label{fig:architectures}
\end{figure}

\section{Decentralized Equalization Architectures}\label{sec:Decentarch}
We start by proposing two feedforward architectures depicted in \fref{fig:architectures}  that enable decentralized equalization and achieve (often significantly) higher spectral efficiency than MRC-based architectures that naturally enable distributed processing~\cite{LETM2014}.

\subsection{System model for decentralized equalization}
We model the input-output relation of a massive MU-MIMO uplink system by $\bmy=\bH\bms_0+\bmn$. 
Here, 
$\bmy\in\complexset^\MR$ is the receive vector and $\MR$ denotes the number of BS antennas, $\bH\in\complexset^{\MR\times\MT}$ is the known MIMO system matrix where each element of~$\bH$ is distributed $\setC\setN(0,1/\MR)$ and $\MT\leq\MR$ denotes the number of users,  $\bms_0\in\setO^\MT$ contains the transmit symbols for each user, $\setO$ is the constellation set (e.g., QPSK), and  $\bmn\in\complexset^\MR$ is i.i.d.\ circularly symmetric complex Gaussian noise with variance $\No$ per complex entry.
%
%
%
We assume an i.i.d. prior $p(\bms_0)=\prod_{\ell=1}^{\MT} p(s_{0\ell})$ and each symbol is distributed as:
\begin{align}\label{eq:prior}
p(s_{0\ell}) =   \textstyle
\frac{1}{|\setO|} \sum_{a\in\setO}  \delta(s_{0\ell} -a),
\end{align}
where $|\setO|$ is the cardinality of $\setO$  and  $\delta(\cdot)$ is the Dirac delta function. The average symbol energy is $\Es=\Ex{}{|s_{0\ell}|^2}$.

As in \cite{LSCCGS2016,LCSGCS2016}, we partition the $\MR$ BS antennas into $\C\geq1$ independent \emph{antenna clusters}. The $c$th cluster is associated with $\MR_c=w_c\MR$ BS antennas so that  $w_c\in(0,1]$ and $\sum_{c=1}^{\C}w_c =1$.
Each cluster contains local radio-frequency~(RF) components and only requires access to local channel state information~(CSI).
Hence, the receive vector for cluster~$c$ can be written as $\bmy_c=\bH_c\bms_{0}+\bmn_c$ with $\bmy_c\in\complexset^{\MR_c}$, $\bH_c\in\complexset^{\MR_c\times\MT}$, and $\bmn_c\in\complexset^{\MR_c}$.
Without loss of generality, we assume the following partitioning: $\bmy=[\bmy_1^\Tran\cdots\bmy_\C^\Tran]^\Tran$ and 
$\bH = [\bH_1^\Tran\cdots\bH_\C^\Tran]^\Tran$.
%


%
%
\subsection{Partially decentralized (\PD) equalization architecture}\label{sec:Decentarch_CE}
The partially decentralized (\PD) equalization architecture is illustrated in \fref{fig:pda} for $C$ clusters.
Each cluster~$c$ independently preprocesses the partial received vector~$\bmy_c$ and channel matrix~$\bH_c$ by computing the partial MRC vector  $\bmymrc_c = \bH_c^\Herm\bmy_c$ and the partial Gram matrix $\bG_c=\bH_c^\Herm \bH_c$.
A feedforward adder tree is used to compute the complete MRC vector and Gram matrix, i.e., computes $\bmymrc=\sum_{c=1}^C\bmymrc_c$ and $\bG=\sum_{c=1}^\C \bG_c$.
Since the MRC output is a sufficient statistic for the transmit signal~\cite{paulraj03}, we perform (linear or non-linear) equalization in a centralized manner and compute a soft symbol $\bmz\in\complexset^\MT$ and variance $\boldsymbol\sigma^2\in\complexset^\MT$ vector, which can be used to either compute hard-output estimates or soft information (e.g., in the form of log-likelihood ratios) of the transmitted bits~\cite{studer2011asic}. 
In \fref{sec:LAMA}, we will analyze the performance of \PD{} equalization for ZF, L-MMSE, and a novel LAMA-based equalization algorithm all of which directly operate on the 
combined 
MRC vector $\bmymrc$ and Gram matrix $\bG$.

%

%
%

%

\subsection{Fully decentralized (\FD) equalization}\label{sec:decent_fda}
The \PD{} architecture requires a summation of both the partial MRC vector and the partial Gram matrices, which involves the transfer and processing of large amounts of data in the adder tree. 
The fully decentralized (\FD) equalization architecture illustrated in \fref{fig:fda} often significantly reduces the overhead of data fusion at the cost of lower performance. 
Specifically, each cluster $c$ independently performs CHEST, preprocessing, \emph{and} equalization, and directly
computes a soft symbol $\bmz_c\in\complexset^\MT$ and variance $\boldsymbol\sigma_c^2\in\complexset^\MT$ vector.
The fusion tree optimally combines the resulting soft symbols $\bmz_c$ and variance $\boldsymbol\sigma_c^2$ vectors in order to generate the final output tuple $\{\bmz,\boldsymbol\sigma^2\}$ used for hard- or soft-output detection; see \fref{sec:FDA} for the details. 

%
%


\section{Partially Decentralized (\PD) Equalization} \label{sec:LAMA}

We start by presenting a decentralized LAMA algorithm suitable for the \PDspace architecture and the associated SE framework.
We then adapt the well-known Tse-Hanly equations \cite{TH1999} to characterize the performance of \PD{} equalization with linear equalization algorithms, such as MRC, ZF, and L-MMSE. 

\subsection{LAMA for \PD{} equalization}\label{sec:LAMA_alg}
The LAMA algorithm \cite{JGMS2015conf} operates on the conventional input-output relation $\bmy=\bH\bms_0+\bmn$. We next propose a novel variant that directly operates on the MRC output $\bmymrc$ and the Gram matrix $\bG$, i.e., the outputs of the fusion tree of the \PD{} architecture shown in \fref{fig:pda}.
Note that since the antenna configuration in massive MU-MIMO systems typically satisfies~\mbox{$\MT\ll\MR$}, the  LAMA-\PD{} algorithm summarized next operates on a lower-dimensional problem while delivering exactly the same results as the original algorithm in  \cite{JGMS2015conf}.


\newtheorem{alg}{Algorithm}
\begin{alg} Initialize $s_\ell=\Exop_S[S]$ for $\ell=1,\ldots,\MT$,  $\phi^1 = \Varop_S[S]$, and $\bmv^1=\mathbf{0}$. Then, for every algorithm iteration $t=1,2,\ldots,$ compute the following steps:\label{alg:lamanew}
\begin{align} \label{eq:LAMA_Gaussian}
\bmz^t &= \bmymrc + (\bI - \bG)\bms^t + \bmv^t \\
\nonumber\bms^{t+1} &= \mathsf{F}(\bmz^t, \No+\beta\phi^t)\\
\nonumber\phi^{t+1} &= \langle\mathsf{G}(\bmz^t, \No+\beta\phi^t)\rangle\\
\nonumber\bmv^{t+1} &= \textstyle \frac{\beta\phi^{t+1}}{\No+\beta\phi^t }(\bmz^t-\bms^t).
\end{align}
The functions $\mathsf{F}(s_\ell,\tau)$ and $\mathsf{G}(s_\ell,\tau)$ correspond to the message mean and variance, respectively, and are computed as follows:
\begin{align}\label{eq:F}
\mathsf{F}(z_\ell,\tau) &=  
\textstyle
\int_{s_\ell} s_\ell f(s_\ell \vert \hat{z}_\ell) \dd s_\ell\\\nonumber
\mathsf{G}(z_\ell,\tau) &= \textstyle\int_{s_\ell} \abs{s_\ell}^2 f(s_\ell \vert \hat{z}_\ell) \dd s_\ell - \abs{\mathsf{F}(z_\ell,\tau)}^2\!.
\end{align}
Here, $f(s\vert {z})$ is the posterior pdf $f(s\vert  z) = \frac{1}{Z}p( z \vert s)p(s)$ with $p( z\vert s)\sim \setC\setN(s,\tau)$, $p(s)$ is given in~\fref{eq:prior}, and $Z$ is a normalization constant. 
\end{alg}

To analyze the performance of LAMA-\PD{} using the SE framework, we need the following definition. 
\begin{defi}[Large-system limit] For a MIMO system with~$\MT$ user antennas and $\MR$ BS antennas, we define the large-system limit by fixing the system ratio $\beta=\MT/\MR$ and letting $\MT\to\infty$.
\end{defi}

We also need the following decoupling property of LAMA.
In the large-system limit and for every iteration $t$, \fref{eq:LAMA_Gaussian} is distributed according to $\setC\setN(\bms_0,\sigma_t^2\bI_\MT)$ \cite{JGMS2015conf}; this shows that LAMA decouples the MIMO system into $\MT$ parallel and independent AWGN channels each with noise variance $\sigma_t^2$.
The SE framework in \fref{thm:SE} with proof in \cite{BM2011} allows us to track the decoupled noise variance $\sigma_t^2$ in each iteration $t$.

\begin{thm} Fix the system ratio $\beta=\MT/\MR$ and the signal prior in \fref{eq:prior}.
%
%
In the large-system limit, the decoupled noise variance $\sigma_t^2$ of LAMA at iteration $t$ is given by the recursion: \label{thm:SE}
\begin{align}\label{eq:SE_recursion}
\sigma_t^2 = \No + \beta\Psi(\sigma_{t-1}^2).
\end{align}
Here, the mean-squared (MSE) function is defined by
\begin{align}\label{eq:SE_MSEpsi}
\Psi(\sigma_{t-1}^2)=\Exop_{S,Z}
\!\Big[
\abs{
  \mathsf{F}(S+\sigma_{t-1}Z,\sigma_{t-1}^2) - S
}^2
\Big],
\end{align}
where $\mathsf{F}$ is given in \fref{eq:F}, $S\sim p(s)$ as in \fref{eq:prior}, $Z\sim\setC\setN(0,1)$, and $\sigma_1^2$ is initialized by $\sigma_1^2=\No+\beta\Varop_S[S]$.
\end{thm}

For $t\to\infty$, the recursion \fref{eq:SE_recursion} 
converges to the fixed-point equation
$\sigma_{\text{PD}}^2 = \No + \beta\Psi(\sigma_{\text{PD}}^2)$.
%
If there are multiple fixed points, then we select the largest $\sigma_{\text{PD}}^2$, which is, in general, a sub-optimal solution \cite{GV2005}.
Since in the large-system limit the MIMO system  is decoupled into 
AWGN channels with noise variance $\sigma_{\text{PD}}^2$ for each user, we will use this fixed-point equation to analyze the achievable rates and error-rate performance of decentralized equalization  in \fref{sec:Results}.

%
%


\subsection{Linear  algorithms for \PD{} equalization}
As for the LAMA-\PD{} algorithm, linear data detectors are also able to operate directly with the MRC output and the Gram matrix.
For MRC equalization, we can directly use the MRC output $\bmymrc$.
%
For ZF and L-MMSE equalization, we first compute a $\MT\times\MT$ filter matrix $\bW = (\bG+\alpha\bI_\MT)^{-1}$, where $\alpha$ is set to $0$ and $\No/\Es$ for ZF and L-MMSE, respectively. The final linear estimate $\bmz$ is then computed by $\bmz = \bW\bmymrc$.

In the large-system limit, the output of MRC, ZF, and L-MMSE-based equalization is also decoupled into AWGN channels~\cite{TH1999} with noise variance $\sigma_\text{PD}^2$ for each user $\MT$. Closed-form expression for the noise variance have been developed by Tse and Hanly in \cite{TH1999}, and are as follows.

\begin{thm}Fix the system ratio $\beta=\MT/\MR$.\label{thm:THeq}
In the large-system limit, the decoupled noise variance $\sigma_{\text{PD}}^2$ for MRC, ZF, and L-MMSE is a fixed-point solution to
$\sigma_{\text{PD}}^2 = \No + \beta\Psi(\sigma_{\text{PD}}^2)$,
%
%
where $\Psi(\sigma^2)$ equals to $\Varop_S[S]$, $\sigma^2$, and $\frac{\Varop_S[S]}{\Varop_S[S]+\sigma^2}\sigma^2$ for MRC, ZF, and L-MMSE equalization, respectively.
\end{thm}

\section{Fully Decentralized (\FD) Equalization}\label{sec:FDA}

We now discuss the algorithm aspects of  the \FD{} architecture shown in \fref{fig:fda} and then analyze its performance.

\subsection{Algorithm procedure for \FD{} architecture}\label{sec:FDA_algo}
Recall from \fref{sec:decent_fda} that each cluster $c$ in the \FD{} architecture independently computes the vectors $\bmz_c$ and $\boldsymbol\sigma_c^2$.
Once equalization for all $C$ clusters is completed, then the vectors $\bmz_c$ and $\boldsymbol\sigma_c^2$ must be fused to compute the output~$\{\bmz,\boldsymbol\sigma^2\}$.
Since the input-output relation from each cluster is decoupled into an AWGN system with i.i.d. noise in the large-system limit, optimal fusion corresponds to computing a weighted sum of $\sum_{c=1}^C\nu_c \bmz_c$
that 
minimizes the output noise variance $\sigma_\text{FD}^2$.
%
%


\subsection{Equalization performance in \FD{} architecture }\label{sec:FDA_LAMA}

The following result characterizes the performance of \FD{} for each cluster $c$; the proof is given in \fref{app:SE_FP_Decentralized}.

\begin{thm} \label{thm:SE_FP_Decentralized}
Let cluster $c$ have a system ratio of $\beta_c = \MT/\MR_c = \beta/w_c$.
In the large-system limit, the input-output relation is decoupled into AWGN channels with noise variance $\bar\sigma_c^2$ given by a solution to the fixed-point equation
%
%
%
$\bar\sigma_c^2 = \textstyle \frac{1}{w_c}\No + \beta_c\Psi(\bar\sigma_c^2)$.
%
Here, $\Psi(\bar\sigma_c^2)$ is the MSE function of the equalizer in cluster $c$.
\end{thm}

Due to the decoupling property in the large-system limit (cf.~\fref{sec:LAMA}), cluster $c$ is decoupled into AWGN channels with noise variance $\bar\sigma_c^2$ for each user.
Hence, fusion relies on $\bmz_c$ and $\bar\sigma_c^2$ for all $C$ clusters and computes a weighted sum that minimizes the final noise variance $\sigma_\text{FD}^2$.
%
%
\fref{lem:optimal_fusion} summarizes the optimal fusion rule; the proof is given in \fref{app:optimal_fusion}.

\begin{lem}
\label{lem:optimal_fusion}
Assume the large-system limit. Let $\bar\sigma_c^2$ be the noise variance for each cluster $c=1,\ldots,C$.
%
Then, the input-output relation of the \FD{} architecture is decoupled into AWGN channels, where the optimal fusion rule is given by
\begin{align}\label{eq:sigma_FD}
 \sigma^2_\text{FD} &= 
 \textstyle 
 \big(\sum_{c=1}^{\C} \frac{1}{\bar\sigma_c^2}\big)^{\!\!-1}
\!=\No + \beta \sum_{c=1}^{\C} \nu_c \Psi(\bar\sigma_c^2)
\end{align}
with $\nu_c = \frac{1}{\bar\sigma_c^2}\big(\sum_{c'=1}^C 1/\bar\sigma_{c'}^2 \big)^{-1}$ for each $c=1,\ldots,\C$.
\end{lem}

We also have the following intuitive result that \FD{} cannot outperform \PD{} equalization; the proof is given in  \fref{app:LAMA_Arch2_Arch1}.
\begin{lem} Let $\No>0$. In the large-system limit, the decoupled noise variances for the  \FD{} and \PD{} architectures satisfy \mbox{$\sigma^2_\text{FD} \geq \sigma_\text{PD}^2$}. Equality holds for $\beta \to 0$ or if MRC is used.\label{lem:LAMA_Arch2_Arch1}
\end{lem}

\section{Numerical Results} \label{sec:Results}

We now investigate the performance of decentralized equalization. 
%
%
We will use an interference-free AWGN channel as the baseline and compare the performance loss (in terms of achievable rates) of \FD{} and \PD{} equalization with MRC, ZF, L-MMSE, and LAMA compared to this baseline.
We define the signal-to-noise ratio (SNR) as $\SNR=\beta\Es/\No$ and the SNR loss as the additional $\Es/\No$ that is required by these equalizers to achieve the same performance as that given by the interference-free AWGN system.
We will assume QPSK modulation and $C=3$ clusters with $w_c=1/C$, $c=1,\ldots,C$.
%
%

\begin{figure*}[tp]
\centering
\subfigure[Fixed rate of $R=1.99$ {[bits/user/channel\,use]}.]{
\includegraphics[width=0.315\textwidth]{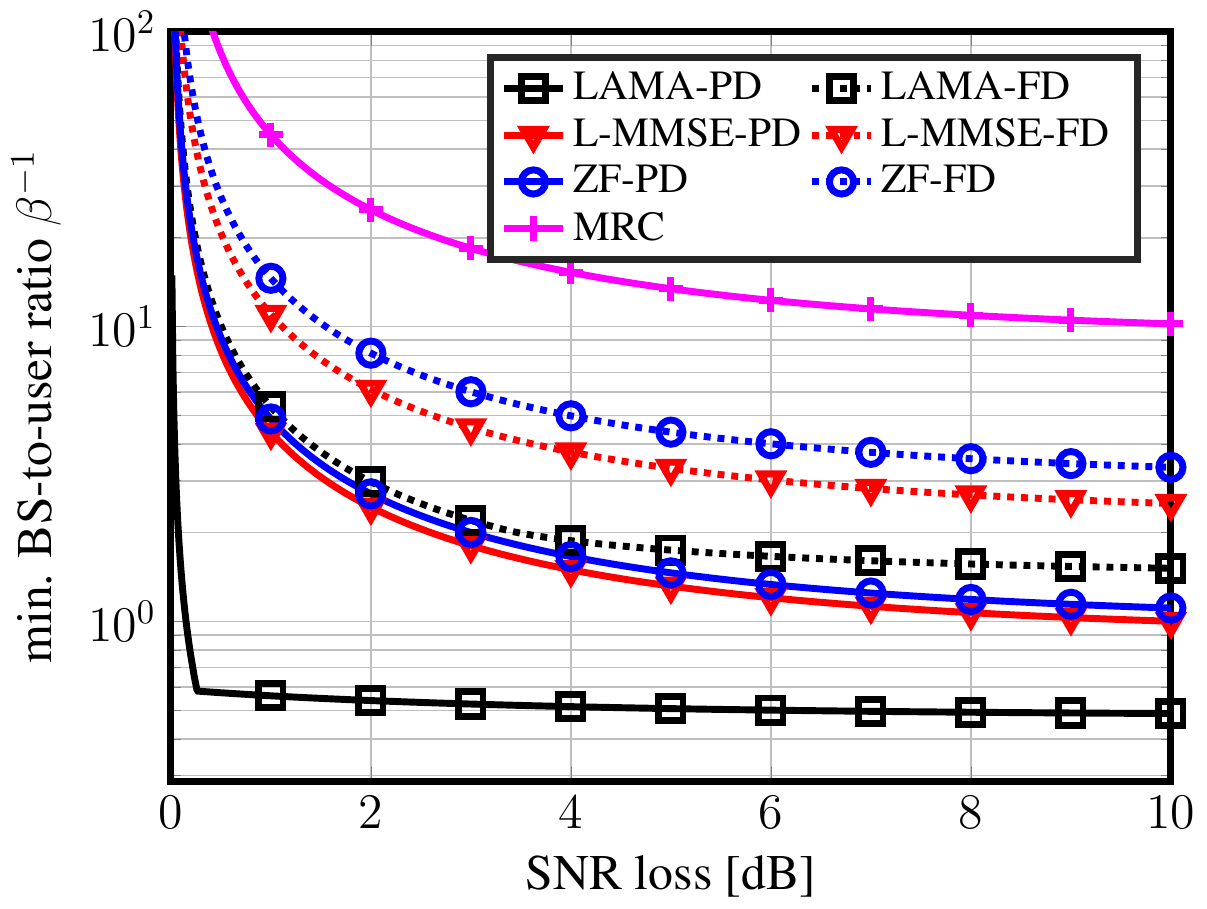}
\label{fig:fig_fixed_rate_1_99}
}
\hspace{-0.1cm}
\subfigure[Fixed  2\,dB SNR loss.]{
\includegraphics[width=0.315\textwidth]{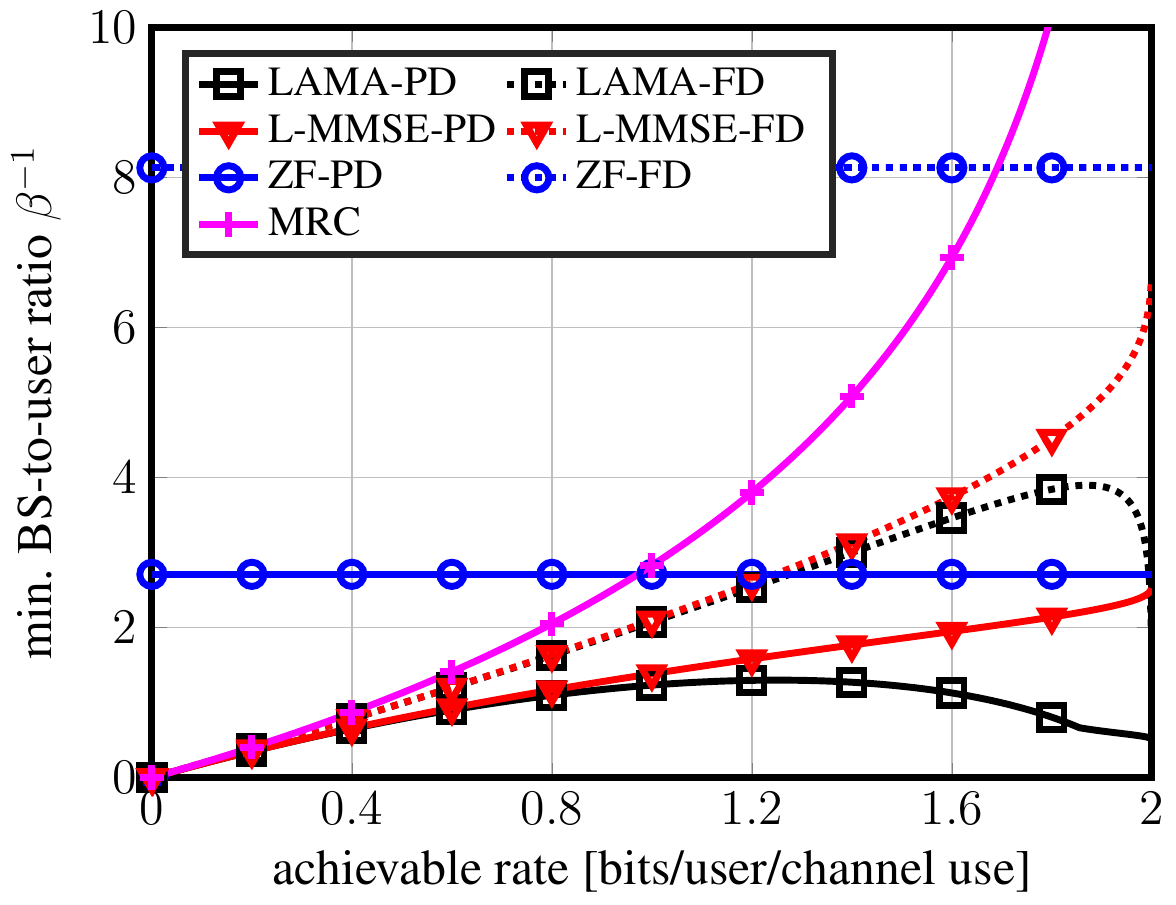}
\label{fig:fig_SNRloss_2}
}
\hspace{-0.1cm}
\subfigure[SER plot for a $96\times16$ and $32\times16$ system.]{
\includegraphics[width=0.315\textwidth]{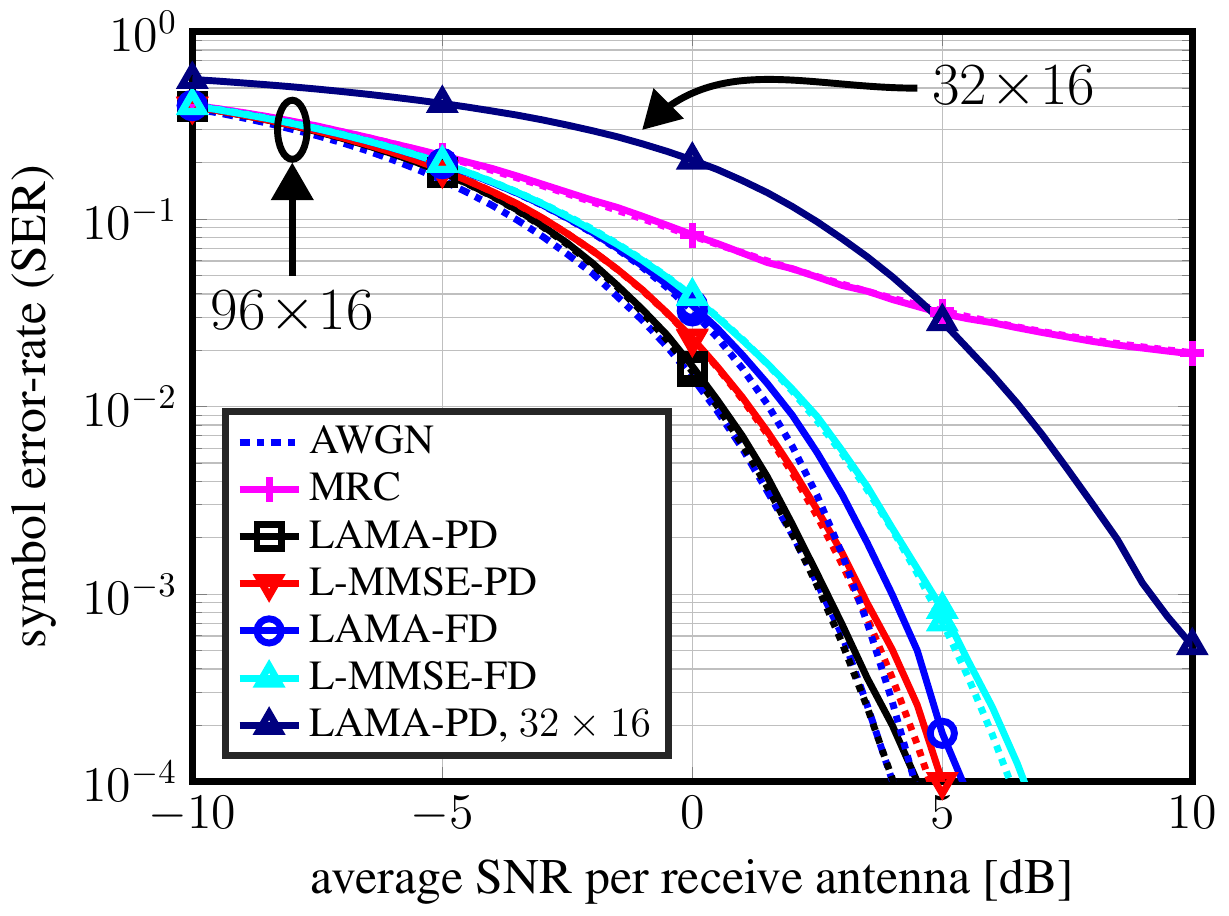}
\label{fig:fig_SER}
}
\caption{
Minimum BS-to-user ratio $\beta^{-1}$ with $C=3$ clusters to achieve (a) a fixed achievable rate of $R=1.99$ and (b) an SNR loss of 2\,dB.
Any $\beta^{-1}$ above the curves are achievable for each system. 
(c) Simulated $96\times16$ MIMO system. 
Our numerical simulations agree with the asymptotic, analytical predictions which are shown with dashed lines. 
FD equalizers for $96\times16$ significantly outperform the error-rate performance of LAMA-PD for a $32\times16$ MIMO system.
%
}
%
\end{figure*}

\subsection{Achievable rate} \label{sec:Results_Rate}

\fref{fig:fixed_rate_1_99} has shown the achievable rates for $\beta^{-1}=6$ for LAMA and L-MMSE for both PD and FD architectures. 
LAMA-FD and L-MMSE-FD suffer only a minimal rate loss compared to the PD counterparts, whereas MRC suffers significant loss in the high-rate regime.
Also, LAMA-FD for $\beta^{-1}=6$ achieves significantly higher rates than that given by LAMA-PD for $\beta^{-1}=2$, which reflects the benefits of data fusion.


We now compute the minimum SNR required to achieve a target rate of $99.5\%$ of full rate (2 bits/user/channel use for QPSK)
for an AWGN channel. We compare the SNR loss for the other equalizers to achieve the same achievable rate for different values of $\beta$. 
\fref{fig:fig_fixed_rate_1_99} shows the results for the different equalization algorithms and decentralized architectures. 
As expected, LAMA-\PD{} significantly outperforms linear algorithms that use \PD{} equalization.
For the \FD{} architecture with an SNR loss of 2\,dB, LAMA-\FD{} achieves the target rate for any $\beta<0.34$. 
LAMA-\FD{} outperforms L-MMSE-\FD{} and suffers only a minimal loss compared to L-MMSE-\PD{}, which requires $\beta<0.16$ and $\beta<0.41$, respectively.
%
%
%
%
This observation implies that LAMA-\FD{} achieves a similar performance as linear equalizers that use the \PD{} architecture while requiring reduced interconnect and chip I/O bandwidth.


\subsection{Minimum required BS-to-user ratio $\beta^{-1}$} \label{sec:Results_MaxBeta}

We fix the SNR loss to 2\,dB and plot the minimum BS-to-user ratio $\beta^{-1}$ for varying achievable rates.
In the low-rate regime, MRC performs similar to all other equalizers, which confirms the well-known fact that MRC is sufficient for massive MU-MIMO in the large-antenna limit \cite{LETM2014}. 
In the high-rate regime, however, MRC requires significantly higher BS-to-user ratios compared to L-MMSE or LAMA-based equalization.
It is interesting to see that the minimum $\beta^{-1}$ decreases for LAMA-\FD{} and LAMA-\PD{} at high rates; this is due to the fact that LAMA in overloaded systems is particularly robust at low and high values of SNR  (see \cite{JGMS2015conf} for more details).

\subsection{Symbol error-rate (SER): analysis vs.\ simulations} \label{sec:Results_SER}

We simulate an uncoded $96\times16$ massive MIMO system and plot the symbol error-rate performance (SER) of LAMA and L-MMSE for the \PD{} and \FD{} architectures with $C=3$ clusters.
We observe that the numerical error-rate simulations (shown with solid lines) closely match the asymptotic predictions (shown with dashed lines for the corresponding color).
We also simulate LAMA-PD for an uncoded $32\times16$ MIMO system as a baseline for comparison with the proposed architectures.

We see that LAMA-\FD{} outperforms L-MMSE-\FD{} and performs close to MMSE-\PD{}.
Furthermore, LAMA-\FD{} performs within 1\,dB of LAMA-\PD{}, which achieves individually-optimal performance in the large-system limit \cite{JGMS2015conf}.
In addition, LAMA-FD with $C=3$ in the $96\times16$ system exhibits significant  performance improvements over LAMA-PD in the $32\times16$ system, which showcases the benefit of the fusion operation in finite-dimensional systems. 
In summary, LAMA-\FD{} delivers near-optimal performance while reducing the interconnect and chip I/O bottlenecks, which demonstrates the efficacy of LAMA and the proposed  feedforward equalization architectures.


\appendices
\section{Proofs}

\subsection{Proof of \fref{thm:SE_FP_Decentralized}}\label{app:SE_FP_Decentralized}
As each entry of the partial matrix channel $\bH_c$ is distributed as $\setC\setN(0,1/\MR)$, we first normalize the system by $1/\sqrt{w_c}$, which amplifies the 
%
%
noise variance by $1/w_c$.
The result follows from Theorems \ref{thm:SE} and \ref{thm:THeq} for LAMA and linear equalization. 


\subsection{Proof of \fref{lem:optimal_fusion}}\label{app:optimal_fusion}

Since in the large-system limit, each input-output relation for $c$th cluster is statistically equivalent to AWGN with noise variance of $\bar\sigma_c^2$, the output of fusion will also be AWGN.
If we define the fusion stage to perform $\bmz = \sum_{c=1}^{\C} \nu_c \bmz_c$, we have
\begin{align*}
\bmz &= 
\textstyle 
\sum_{c=1}^{\C} \nu_c \bmz_c = \sum_{c=1}^{\C} \nu_c\bms_0 + \sum_{c=1}^{\C} \nu_c \bar\sigma_c \bmn_c 
\stackrel{(a)}{=} \bms_0 + \sigma_\text{FD}\bar\bmn,
\end{align*}
where $(a)$ follows from $\sum_{c=1}^{\C} \nu_c=1$ and $\bmn_c\sim\setC\setN(0,\bI_{\MT})$ are independent for all $c=1,\ldots,\C$.
Here, we have that 
$\sigma^2_\text{FD} = \sum_{c=1}^{\C} \nu_c^2 \bar\sigma_c^2$ and $\bar\bmn\sim\setC\setN(0,\bI_{\MT})$.

We minimize 
 the noise variance 
$\sigma^2_\text{FD}$ subject to the constraint $\sum_{c=1}^{\C} \nu_c=1$, which
%
gives $\nu_c = \frac{1}{\bar\sigma_c^2}\big(\sum_{c=1}^C 1/\bar\sigma_c^2 \big)^{-1}$ for all 
$c=1,\ldots,\C$.
Obtaining the first expression in \fref{eq:sigma_FD} is straightforward; the second expression is obtained 
as follows:
\begin{align*}
& 
\textstyle 
\beta\sum_{c=1}^{\C} \nu_c \Psi(\bar\sigma_c^2) 
= 
\left(\sum_{c=1}^{\C}\frac{1}{\bar\sigma_c^2}\right)^{\!\!-1}\!
\sum_{c=1}^{\C} \frac{\beta\Psi(\bar\sigma_c^2)}{\bar\sigma_c^2}
\\&= \textstyle
\left(\sum_{c=1}^{\C}\frac{1}{\bar\sigma_c^2}\right)^{\!\!-1}\!
\sum_{c=1}^{\C}\left(w_c - \frac{\No}{\bar\sigma_c^2}
\right)
=
\left(\sum_{c=1}^{\C}\frac{1}{\bar\sigma_c^2}\right)^{\!\!-1}\!\!\! - \No.
\end{align*}

\subsection{Proof of \fref{lem:LAMA_Arch2_Arch1}}\label{app:LAMA_Arch2_Arch1}

%

We first show when equality holds. 
The case for $\beta \to 0$ is trivial because $\bar\sigma_c^2=\sigma_\text{FD}^2=\sigma_\text{PD}^2 = \No$.
For MRC, we have $\sigma_\text{FD}^2 = \No+\beta\sum_{c=1}^{\C} \nu_c \Varop_S[S] = \No+\beta\Varop_S[S] = \sigma_\text{PD}^2$.

Let us now assume that $\beta>0$.
We show $\bar\sigma_c^2 > \sigma_\text{PD}^2$ by re-writing the fixed-point solutions as \cite{Maleki2010phd}:
$\bar\sigma_c^2 = \sup\{ \sigma^2: \No+\beta\Psi(\sigma^2)\geq w_c\sigma^2\}$ 
and 
$\sigma_\text{PD}^2 = \sup\{ \sigma^2: \No+\beta\Psi(\sigma^2)\geq \sigma^2\}$.
Note that $\No>0$, so both $\bar\sigma_c^2$ and $\sigma_\text{PD}^2$ are strictly positive.
It is easy to see that $\sigma_\text{PD}^2 \neq \bar\sigma_c^2$ because $\sigma_\text{PD}^2  = \No+\beta\Psi(\sigma_\text{PD}^2 ) > w_c\sigma_\text{PD}^2$.
%
%
Since $\Psi(\sigma^2 ) \to \Varop_S[S]$ as $\sigma^2\to\infty$ and $\Psi(\sigma^2)$ is continuous \cite{GWSS2011}, there exists a $\bar\sigma_c^2>\sigma_\text{PD}^2$ that satisfies $\No+\beta\Psi(\bar\sigma_c^2)=w_c\bar\sigma_c^2$ by the intermediate value theorem.
%

Finally, we use \cite[Prop. 9]{GWSS2011} to see that $\Psi(\sigma^2)$ is strictly increasing for $\sigma^2>0$ for LAMA. For ZF and MMSE, this also holds by computing $\dd \Psi(\sigma^2)/\dd\sigma^2 > 0$. 
Thus, the result $\sigma_\text{FD}^2 > \sigma_\text{PD}^2$ follows directly from \fref{lem:optimal_fusion} since
\begin{align*}
\sigma^2_\text{FD}  =\No \!+\! \beta \sum_{c=1}^{\C} \nu_c \Psi(\bar\sigma_c^2) > \No \!+\! \beta \sum_{c=1}^{\C} \nu_c \Psi(\sigma_\text{PD}^2) 
= \sigma_\text{PD}^2.
\end{align*}

\bibliographystyle{IEEEtran}
\bibliography{bib/VIPabbrv,bib/confs-jrnls,bib/publishers,bib/VIP_170507}


\end{document}